\documentclass[12pt]{article}
\usepackage{geometry}
\usepackage{amssymb}
\usepackage{amsmath}
\usepackage{MnSymbol}
\usepackage{hyperref}
\usepackage{braket}

\numberwithin{equation}{section}

\newcommand{\be}{\begin{equation}}
\newcommand{\ee}{\end{equation}}
\newcommand{\bea}{\begin{eqnarray}}
\newcommand{\eea}{\end{eqnarray}}

\newcommand{\A}{\mathcal{A}}
\newcommand{\C}{\mathbb{C}}

\newcommand{\R}{\mathbb{R}}

\newcommand{\Z}{\mathbb{Z}}
\newcommand{\Or}{\mathcal{O}}

\newcommand{\im}{\text{Im}}

\begin{document}
		\begin{titlepage}

		\begin{center}

			\vskip 2 cm
			{\Large \bf The Genus Two Free Boson in Arakelov Geometry}\\
			\vskip 1.25 cm {Thomas Vandermeulen\footnote{email address: tvandermeulen@albany.edu}}\\
			
			{\vskip 0.5cm \it Department of Physics, University at Albany, \\ Albany, NY 12222, USA \\}
			
		\end{center}
		\vskip 2 cm
		
		\begin{abstract}
			\baselineskip=18pt
			Using Arakelov geometry, we compute the partition function of the noncompact free boson at genus two.  We begin by compiling a list of modular invariants which appear in the Arakelov theory of Riemann surfaces.  Using these quantities, we express the genus two partition function as a product of modular forms, as in the well-known genus one case.  We check that our result has the expected obstruction to holomorphic factorization and behavior under degeneration.
			
		\end{abstract}
		
	\end{titlepage}

\section{Introduction}

The free boson, or free scalar field $\varphi$, is one of the simplest examples of a Conformal Field Theory (CFT).  Here we focus on the two-dimensional case, where $\varphi$ can be taken to be a function of the complex variables $z,\bar{z}$.  Due to conformal invariance, $z,\bar{z}$ can be taken as living on some compact Riemann surface $X_g$ of genus $g$, which shows up in the theory's action as
\be
S[\varphi]=\int_{X_g}d^2z\partial\varphi\bar{\partial}\varphi.
\ee
The partition function is given by a path integral of the form
\be
\label{zgpathint}
Z_g=\int \mathcal{D}\varphi e^{-S[\varphi]}
\ee
which, given that $S$ is quadratic in $\partial\varphi$, takes the form of an infinite-dimensional gaussian integral.  In fact we can exploit this gaussian form to evaluate (\ref{zgpathint}) as
\be
\label{zgdet}
Z_g=\left[\frac{(\det{\Delta_0})_{X_g}}{V_{X_g}}\right]^{-\frac{1}{2}},
\ee
where $(\det{\Delta_0})_{X_0}$ is the zeta function-regularized determinant of the scalar laplacian on $X_g$ and $V_{X_g}$ is the surface's volume.

For low $g$, simplifications relevant to CFT of (\ref{zgdet}) are well-known.  On a sphere we have no moduli for $Z_0$ to depend on, so we can choose normalization such that it is simply equal to 1.  $Z_1$, the torus partition function for the free boson, is a little more interesting in that it can depend on the torus' complex structure constant $\tau$.  It takes the form
\be
\label{z1}
Z_1(\tau)=\frac{|\eta(\tau)|^{-2}}{\sqrt{\im\tau}}.
\ee

Moving to $g>1$, we begin to find more indirect expressions.  As (\ref{zgdet}) was stated to be zeta-function regularized, we can formally define a zeta function from the eigenvalues of $\Delta_0$ and write $Z_g$ in terms of it \cite{DHokerPhongRegularization}.  A more usual presentation for $Z_g$ is given in terms of theta functions, Green's functions and products of determinants of one-forms \cite{AMNVB,TwoLoopIV,VV}.  Such formulas stem from the relation of $Z_g$ to Faltings' delta invariant \cite{Faltings} -- for a taste of what the precise relation looks like, one could refer ahead to the definition of the delta invariant (\ref{delta}) and its relation (\ref{zdrelation}) to the partition function.

Ideally we would like, for any genus $g$, to have a nice, closed-form expression analogous to (\ref{z1}) for the partition function $Z_g$, but to avoid being overly ambitious we will start with $Z_2$.  Such simplifications have been sought before.  Notably, \cite{Tuite} begins from a similar premise, and uses sewing procedures to derive a series form for $Z_2$.  This paper aims to attack the same problem from a different point of view.  We take advantage of the aforementioned relation between $Z_2$ and Faltings' delta invariant.  This, along with work from the mathematical literature expressing the delta invariant in terms of more elementary modular functions, will allow us to find simple expressions for $Z_2$.

It may appear that what we are searching for is dangerously specific, in that it would apply only to one simple CFT.  However, the free boson partition function has a much more universal relevance.  Free bosons (and fermions, through the fermionization process) serve as building  blocks for more complicated CFTs, via such procedures as orbifolding and deformation \cite{Ginsparg}.  Additionally, the noncompact free boson partition function appears in the characters of the minimal models \cite{MathurSen}.  Moreover, pure CFT at higher genus has undergone a renewed interest due to the modular bootstrap program \cite{Keller,CMM,CCY}.  Having an explicit expression in terms of modular forms for one of the fundamental objects in genus two CFT should be quite valuable to these investigations.

Section \ref{AlgGeom} will introduce the requisite preliminary notions of algebraic geometry.  After establishing conventions for basic quantities we will be prepared to give the definitions relevant to the Arakelov geometry of Riemann surfaces.  This culminates in the enumeration of a variety of modular functions associated to these surfaces.  Section \ref{Degen} continues to develop machinery -- namely the separating degeneration of a genus two Riemann surface.  Knowing the behavior of the period matrix allows us to expand more complex quantities in a degeneration series.  The results are brought together in section \ref{CFT}, where we will derive explicit expressions for $Z_2$.  Applying the degeneration procedure allows us to check our findings against pure CFT calculations.  Section \ref{conclusion} will summarize our efforts and give a look at future prospects.

\section{Algebraic Geometry Preliminaries}
\label{AlgGeom}

In order to study quantities defined on the moduli space of Riemann surfaces, we need to develop the basic notions of algebraic geometry on such surfaces.  The focus of this paper will be on genus two surfaces, and in particular degenerating families of genus two surfaces.

\subsection{One-Forms, Period Matrix and Theta Functions}

On a Riemann surface $X_g$ of genus $g$ we have dim$H^1(X;\R)=2g$, meaning that a basis for cohomology consists of a choice of $g$ holomorphic and $g$ antiholomorphic one-forms.  We call these $\omega_i$ ($\bar{\omega}_i$), and will tend to leave implicit the accompanying differential when writing them.  Along with the one-forms we have a corresponding basis of $2g$ cycles in homology.  We divide these into $g$ $a$-cycles $a_i$ and $g$ $b$-cycles $b_i$, normalized under the intersection product $\circ$ as
\be
\label{homologybasis}
a_i\circ b_j=\delta_{ij},\hspace{.5cm}a_i\circ a_j=b_i\circ b_j=0.
\ee

The pairing between homology and cohomology is given by the periods of the one-forms over these cycles.  We normalize these quantities such that
\be
\label{omegaperiods}
\int_{a_j}\omega_i=\delta_{ij}\hspace{1cm}\int_{b_j}\omega_i=\tau_{ij},
\ee
where the matrix $\tau_{ij}$ is called the period matrix.  It is complex with positive-definite imaginary part, and its entries are taken to characterize the moduli of our surface.\footnote{At genera above three we would need to impose additional relations on the period matrix as the number of free parameters in $\tau$ exceeds the number of moduli.  Since our explicit calculations will not go beyond genus two we do not worry about this.}

We define the basic Siegel theta function at genus $g$ by
\be
\label{thetafunc}
\theta(z|\tau)=\sum_{x\in\Z^g}\exp{\left[\pi ix\cdot\tau\cdot x+2\pi ix\cdot z\right]}.
\ee
It takes as its arguments a $g$-vector $z$ and a $g\times g$ matrix $\tau$.  In order to associate this object with a Riemann surface, we let $\tau$ be that surface's period matrix.  For the remaining argument, we note that from any point $z\in X$ we can construct a $g$-vector as
\be
\int_{z_0}^z\omega_i.
\ee
Such a procedure takes a point on $X$ to its Jacobian, $\C^g/(\Z^g+\tau\Z^g)$, and is known as the Abel map.  There is a dependence on the choice of basepoint $z_0$ which needs to be resolved on a case-by-case basis.  Whenever we speak of a theta function depending on coordinates on a Riemann surface, we are implicitly including the Abel map.


Further, one defines a theta function with characteristics as
\be
\label{thetacharac}
\theta\genfrac[]{0pt}{0}{\delta}{\epsilon}(z|\tau)=\sum_{x\in\Z^g}\exp{\left[\pi i\left(x+\delta\right)\cdot\tau\cdot\left(x+\delta\right)+2\pi i\left(x+\delta\right)\cdot\left(z+\epsilon\right)\right]}.
\ee
One usually takes half-integer characteristics $(\delta,\epsilon)\in(\frac{1}{2}\Z_2^g)^2$.  The choice of 0 or 1/2 for each component of $\delta$ and $\epsilon$ can be viewed as a periodicity assignment along the $a$ and $b$ cycles of the Riemann surface, and because of this are also known as spin structures.  These functions are alternatively written in terms of the basic theta function with a shifted first argument as
\be
\label{charrelation}
\theta\genfrac[]{0pt}{0}{\delta}{\epsilon}(z|\tau)=\exp{\left[2\pi i\left(\frac{1}{2}\delta\cdot\tau\cdot\delta+\delta\cdot z+\delta\cdot\epsilon\right)\right]}\theta\left(z+\epsilon+\tau\delta|\tau\right).
\ee
In particular, (\ref{charrelation}) can be used to rewrite ratios of theta functions with characteristics as ratios of shifted theta functions.  Theta functions (with characteristics) evaluated at $z=0$ are known as theta constants.  

Under $z\to-z$, theta functions with half-integer characteristics acquire a phase of $\exp{[4\pi i(\delta\cdot\epsilon)]}$.  Accordingly, then, the function will be even or odd in $z$ depending on whether $4(\delta\cdot\epsilon)$ mod 1 is 0 or 1, respectively.  In the same way, we call a characteristic even or odd based on $4(\delta\cdot\epsilon)$ mod 1.  An easy consequence of this parity is that theta constants with odd characteristics vanish identically in $\tau$.

Theta functions are \textit{almost} periodic with respect to the Jacobian lattice $\Z^g+\tau\Z^g$, having the quasi-periodicity property
\be
\theta(z+\beta\cdot\tau+\alpha|\tau)=\exp{[-\pi i\beta\cdot\tau\cdot\beta-2\pi i\beta\cdot z]}\theta(z|\tau)
\ee
with $(\alpha,\beta)\in (\Z^g)^2$.  We note, then, that the  Peterson norm, defined as\footnote{Throughout this paper we will define double bar norms of various objects.  Though the exact meaning will vary between objects, the procedure should be understood as
	\begin{itemize}
		\item If the object depends on coordinates on the Riemann surface, make it periodic with respect to the Jacobian lattice.
		\item If the object has nonzero modular weight, insert factors of $\det{\im\tau}$ to make it weight 0.
		\item Possibly adjust normalization with a multiplicative constant.
	\end{itemize}
Note that the factors required to ensure periodicity (as in (\ref{thetanorm})) will mix holomorphic and antiholomorphic terms, so in general this comes at the cost of holomorphic factorization.
}
\be
\label{thetanorm}
||\theta(z|\tau)||=(\det{\im\tau})^{1/4}\exp{[-\pi \im z\cdot\im\tau^{-1}\cdot\im z]}|\theta(z|\tau)|
\ee
is truly periodic under these shifts.  We show one further useful property of this norm.  Using (\ref{charrelation}), we write
\be
\label{charshift1}
|\theta\genfrac[]{0pt}{0}{\delta}{\epsilon}(z|\tau)|=\exp{[-\pi\delta\cdot\im \tau\cdot\delta-2\pi\delta\cdot\im z]}|\theta(z+\delta\cdot z+\epsilon|\tau)|.
\ee
One readily verifies that (again $(\alpha,\beta)\in (\Z^g)^2$)
\be
\label{thetacharnorm}
|\theta\genfrac[]{0pt}{0}{\delta}{\epsilon}(z+\beta\cdot\tau+\alpha|\tau)|=\exp{[\pi\beta\cdot\im \tau\cdot\beta+2\pi\beta\cdot\im z]}|\theta\genfrac[]{0pt}{0}{\delta}{\epsilon}(z|\tau)|
\ee
which tells us that (\ref{thetanorm}) remains valid for theta functions with characteristics.  Multiplying both sides of (\ref{charshift1}) by $(\det{\im\tau})^{1/4}\exp{[-\pi \im z\cdot\im\tau^{-1}\cdot\im z]}$ then yields
\be
\label{thetanormcharswap}
||\theta\genfrac[]{0pt}{0}{\delta}{\epsilon}(z|\tau)||=||\theta(z+\delta\cdot\tau+\epsilon|\tau)||,
\ee
so that within norms we are allowed to interchange characters with shifts by their corresponding half-periods.

Let $\Delta$ correspond to an odd characteristic for which $\theta_\Delta(z)$ does not vanish identically.  Define the holomorphic differential
\be
\omega_\Delta(z)=\sum_{i=1}^g\partial_{z_i}\theta_\Delta(0|\tau)\omega_i(z)
\ee
where $z$ is some local coordinate on $X$.  It is straightforward to show that $\omega_\Delta(z)$ vanishes where $\theta_\Delta(z)$ does; in fact, $\omega_\Delta$ has double zeroes at those points.  It is then sensible to take a square root of $\omega_\Delta$,
\be
\label{spinor}
h_\Delta(z)=\sqrt{\sum_{i=1}^g\partial_{z_i}\theta_\Delta(0|\tau)\omega_i(z)},
\ee
which is a half-differential or spinor, having now single zeroes commensurate with $\theta_\Delta$.

We are now prepared to write down the prime form,
\be
\label{primeform}
E(z,w)\equiv\frac{\theta_\Delta(z-w|\tau)}{h_\Delta(z)h_\Delta(w)},
\ee
where the notation indicates the nonobvious fact that $E$ does not depend on the choice of $\Delta$ (provided it's odd and nonsingular).  The prime form is the generalization of the function $z-w$ to arbitrary Riemann surfaces, and is useful for constructing meromorphic functions on those surfaces.

\subsection{Modular Transformations}

A choice of canonical homology basis is not unique.  From (\ref{homologybasis}), we see that the intersection products are of the form
\be
\label{symplecticform}
\left(\begin{matrix}a\circ a & a\circ b \\ b\circ a & b\circ b\end{matrix}\right)=\left(\begin{matrix}0 & I \\ -I & 0\end{matrix}\right).
\ee
Any change of basis leaving the right-hand side of (\ref{symplecticform}) invariant produces a new canonical basis.  Such a transformation is given by an element of Sp$(2g;\Z)$ -- these are the modular transformations of our surface, and can be written as a matrix with integer entries of the form
\be
\left(\begin{matrix}A & B \\ C & D\end{matrix}\right):\hspace{.2cm}AB^T-BA^T=CD^T-DC^T=0,\hspace{.2cm}AD^T-BC^T=1.
\ee
Under such a transformation, the one-forms and period matrix change as
\be
\omega_i\to\omega_j(C\tau+D)^{-1}_{ji} \hspace{.5cm} \tau\to(A\tau+B)(C\tau+D)^{-1}.
\ee

General modular functions are classified by their weights: an object of weight $(n,m)$ acquires a factor of $(\det{[C\tau+D]})^n(\det{[C\bar{\tau}+D]})^m$ under modular transformations.\\

\subsection{Specialization to Genus One}

All compact, orientable Riemann surfaces of genus one are equivalent to the torus, which can be described as the lattice $\C/(\Z+\tau\Z)$.  Such a surface is then equivalent to its own Jacobian.  Here $\tau\in\mathbb{H}$ is a single complex number, often called the complex structure constant of the torus.  Due to the simplicity of this case, we are able to write explicit expressions for almost all relevant quantities.

\subsubsection{Basic Quantities}

At this genus there are four half-integer characteristics, three even and one odd.  Their associated theta functions are written as:
\be
\theta_1(z|\tau)\equiv\theta\left[\begin{matrix}\frac{1}{2} \\ \frac{1}{2}\end{matrix}\right](z|\tau),\hspace{.25cm}\theta_2(z|\tau)\equiv\theta\left[\begin{matrix}\frac{1}{2} \\ 0\end{matrix}\right](z|\tau),\hspace{.25cm}\theta_3(z|\tau)\equiv\theta\left[\begin{matrix}0 \\ 0\end{matrix}\right](z|\tau),\hspace{.25cm}\theta_4(z|\tau)\equiv\theta\left[\begin{matrix}0\\ \frac{1}{2}\end{matrix}\right](z|\tau). 
\ee
Among the many identities satisfied by these functions are those relating the theta constants to the dedekind eta function:
\be
\label{thetaetarelation}
\theta_2(0|\tau)\theta_3(0|\tau)\theta_4(0|\tau)=-\frac{1}{\pi}\frac{\partial}{\partial z}\theta_1(0|\tau)=2\eta^3(\tau).
\ee

We can take the coefficient of the single one-form on the torus to simply by unity.  This allows us to write out the spinor (\ref{spinor}) explicitly.  We have, using (\ref{thetaetarelation}),
\be
h_\Delta(z)=\sqrt{\theta'_1(0|\tau)}=i\sqrt{2\pi}\eta^{3/2}(\tau).
\ee
We can, as well, write the torus prime form as
\be
\label{torusprimeform}
E(z,w)=\frac{\theta_1(z-w|\tau)}{\theta'_1(0|\tau)}=\frac{\theta_1(w-z|\tau)}{2\pi\eta^3(\tau)}
\ee

\subsubsection{Integral Calculations}
\label{integralcalc}

Later we will find it useful to perform integrals over tori, so we outline a typical calculation here and prove some useful results.  Following \cite{dejongthesis}, we choose a fundamental domain in $\C$ for our torus given by $z=\alpha\tau+\beta$ with $\alpha\in[-1/2,1/2]$ and $\beta\in[0,1]$. (\ref{mu}) becomes, in the case of a torus,
\be
\label{torusmu}
\mu =\frac{idz\wedge d\bar{z}}{2\im \tau}=d\alpha d\beta.
\ee
A typical quantity to integrate is $\log{||\theta(z|\tau)||}$, which we now demonstrate.  Using (\ref{thetanorm}), this integral will be written as
\be
\int^{1/2}_{-1/2}d\alpha\int^1_0 d\beta \left[-\pi\frac{(\im z)^2}{\im\tau}+\log{|\theta(z|\tau)|}+\frac{1}{4}\log{\im\tau}\right].
\ee
The first term straightforwardly evaluates to $-\frac{\pi}{12}\im\tau$.  For the second, we write the theta function in its product form
\be
\label{thetaprod}
\theta(z|\tau)=\prod_{m=1}^\infty (1-\exp{(2\pi i m\tau)})(1+\exp{(\pi i(2m-1)\tau+2\pi iz)})(1+\exp{(\pi i(2m-1)\tau-2\pi iz)}).
\ee
We note that, after taking a log, the second and third terms in the product can be expanded as
\be
\sum_{k=1}^\infty \frac{(-1)^{k-1}\exp{[\pi ik(2m-1)\tau\pm2\pi ikz]}}{k}.
\ee
Due to the periodicity of these terms in $z$, the integration in $\beta$ kills them off (this applies holomorphically and antiholomorphically), so we are left with
\be
\label{intthetatorus}
\int_{T_1}\log{||\theta(z|\tau)||}\cdot\mu=-\frac{\pi}{12}\im\tau+\log{\prod_{m=1}^\infty|1-\exp{(2\pi im\tau)}|}+\frac{1}{4}\log{\im\tau}=\log{||\eta(\tau)||}.
\ee

We obtain an additional result by dividing (\ref{intthetatorus}) by $(\im\tau)^{1/4}$ and differentiating in $\tau$:
\be
\label{dintthetatorus}
\int\left[\partial_\tau\log{\theta(\alpha\tau+\beta|\tau)}+\frac{\pi i}{2}\alpha^2\right]d\alpha d\beta=\partial_\tau\log{|\eta(\tau)|}=\frac{1}{2}\partial\log{\eta(\tau)}.
\ee

\subsection{Arakelov Geometry and Modular Invariants at Genus Two}
\label{Arakelov}

In the context of the algebraic geometry of Riemann surfaces, Arakelov geometry involves a particular choice of metric.  We begin by defining a (1,1)-form on our surface $X$ as follows \cite{Wentworth}:
\be
\label{mu}
\mu=\frac{i}{2g}\sum_{i,j}(\text{Im}\tau)^{-1}_{ij}\text{ }\omega_i\wedge\bar{\omega}_j=\mu_{z\bar{z}}dz\wedge d\bar{z}.
\ee
Since $\int_X\omega_i\wedge\bar{\omega}_j=-2i\im\tau_{ij}$, we see that the integral of $\mu$ over our surface is normalized to 1.  On the Jacobian we can define a similar form
\be
\label{nu}
\nu=\frac{i}{2}\sum_{i,j}(\text{Im}\tau)_{ij}^{-1}dZ_i\wedge d\bar{Z}_j.
\ee
The Abel map $\A:X\to J(X)$ gives us, under the usual pullback of forms, the relation $g\mu=\A^{*}\nu$.  The Haar measure on the Jacobian (regarded as the topological group $\C^g/(\Z^g+\tau \Z^g)$) is given by $\nu^g/g!$.

From $\mu$ we can define the Arakelov Green's function as the unique function $G(z,w)$ satisfying \cite{Arakelov}
\begin{itemize}
	\item $G(z,w)=G(w,z)$. \hfill\refstepcounter{equation}\label{akgreensymm}(\theequation)
	\item $\int_{X} \log{G(z,w)}\cdot\mu(w)=0$\hfill\refstepcounter{equation}\label{akgreenrel}(\theequation)
	\item G has a zero of order one for $z=w$.
	\item For $z\neq w$, $\partial_z\bar{\partial}_zG(z,w)=\pi i\mu_{z\bar{z}}$.
\end{itemize}
Further, we can define the Arakelov metric $2g_{z\bar{z}}dzd\bar{z}$ by requiring its curvature to be proportional to $\mu$:
\be
\label{arakelovmetric}
\partial_z\partial_{\bar{z}}\log{g_{z\bar{z}}}=4\pi i(g-1)\mu_{z\bar{z}}.
\ee
The Arakelov metric, Arakelov Green's function and (the norm of) the prime form on $X$ are related by \cite{DHokerPhongMandelstam}
\be
\label{GEgRelation}
2\log{G(z,w)}=\log{||E(z,w)||^2}+\frac{1}{2}\log{g_{z\bar{z}}}+\frac{1}{2}\log{g_{w\bar{w}}}+\log{\det{\im\tau}}.
\ee
With these ingredients we can now define a number of modular invariant quantities.  Following the mathematical literature we write them as functions of the surface -- the goal will eventually be to reduce all of our expressions to functions of the period matrix.
\begin{itemize}
\item We begin with the delta invariant $\delta(X)$ of Faltings \cite{Faltings}, which will be of great use for its relation to the determinant of the laplacian.  Thus we mainly use $\delta$ as an intermediary to write $\det{\Delta}$ in terms of other, more simple invariants.  Though we will not use it, we can give an independent definition of $\delta$ as \cite{Wentworth}
\be
\label{delta}
\exp{(-\delta(X)/8)}=\frac{||\theta(\sum_{i=1}^gz_i-w-K|\tau)||}{||\det{ \omega_i(z_j)||}}\frac{\prod_{i<j}^gG(z_i,z_j)}{\prod_{i=1}^gG(z_i,w)}.
\ee
Here the norm $||\det{\omega_i(z_j)}||$ is defined by
\be
||\det{\omega_i(z_j)}||^2=\prod_{k=1}^g(g_{z_k\bar{z}_k})^{-1}|\det{\omega_i(z_j)}|^2,
\ee
and $K$ is a quantity known as the Riemann vector, the form of which is unimportant for our calculations.  Though non-obvious, (\ref{delta}) is independent of the $g+1$ points $z_i,w$.

\item Another invariant,$A(X)$ appears in a formula, due to Bost, for the Arakelov Green's function \cite{Bost},
\be
\log{G(q,w)}=\int_{\Theta+\A(q-w)}\log{||\theta(Z|\tau)||}\cdot\frac{\nu^{g-1}}{g!}+A(X),
\ee
where integration is taken over the theta divisor, $\Theta=\{Z\in J(X)|\theta(Z|\tau)=0\}$.  We explicitly include the Abel map on $q-w$ to emphasize that the integral is taken over a shifted region in the Jacobian.  Noting (\ref{akgreensymm}) and (\ref{akgreenrel}), one readily calculates $A$ as
\be
A(X)=-\int_{X}\mu(w)\int_{\Theta+\A(q-w)}\log{||\theta(Z|\tau)||}\cdot\frac{\nu^{g-1}}{g!}.
\ee
Translation invariance of $\nu$ guarantees that $A$ does not depend on the choice of $q$. We can recast $A$ purely in terms of integration over the Riemann surface instead of its Jacobian.  Specifically, for $\Theta$ we have the relation
\be
\label{thetadivisor}
\theta\left(\A(\sum_{i=1}^{g-1} z_i)-K\bigg|\tau\right)=0
\ee
for all $z_i\in X$, and this parameterizes the entirety of $\Theta$ \cite{Bobenko}.  This allows us to pull back the inner integral into one over $X$.  Choosing $q$ such that $\A(q)-K$ gives a half period $\Delta$, we arrive at 
\be
A(X_2)=-\int_{X_2^2}\log{||\theta_\Delta(\A(z-w)|\tau)||}\cdot\mu(z)\mu(w)
\ee
where the result does not depend on the specific choice of $\Delta$.

\item Let $h_\Delta$ be a spinor corresponding to the odd, nonsingular spin structure $\Delta$.  Define
\be
\label{C}
\log{C(X)}=\int_{X}\log{[(\det{\text{Im}\tau})^{1/8}|h_\Delta(z)|]}\cdot\mu(z).
\ee
$C$ is independent of $\Delta$ (see \cite{DHokerGreen}, Appendix A).

\item Another invariant, due to De Jong \cite{DeJong}, is
\be
\label{S}
\log{S(X)}=-\int_{X}\log{||\theta(gz-w|\tau)||}\cdot\mu(z).
\ee
Translation invariance on $J(X)$ ends up guaranteeing that the result is $w$-independent.

\item There is then an invariant $T(X)$ which relates $S(X)$ to $\delta(X)$:
\be
\label{dstrelation}
\exp{(\delta(X)/4)}=S(X)^{-\frac{g-1}{g^2}}T(X).
\ee
For hyperelliptic surfaces (which include all genus two surfaces) $T$ is proportional to the discriminant modular form, which in turn can be expressed as a product of theta functions.

\item We have a quantity similar to $S$, but defined by integration over the Jacobian of $X$ \cite{Wilms}:
\be
\label{H}
\log{H(X)}=\int_{J(X)}\log{||\theta(Z|\tau)||}\cdot\frac{\nu^g}{g!}.
\ee

\item  Specializing to genus two, we can define a modular form by extracting the factor of $\det{\im\tau}$ implicit in the normed theta function in (\ref{H}).   Following the notation of \cite{DHokerGreen}, we call the (square of) the result $\Phi$:
\be
\label{PhiHrel}
\Phi(\tau)=\frac{H^2(X_2)}{\sqrt{\det{\im\tau}}}=\exp{\left[\int_{J(X)}\left[\log{|\theta(Z|\tau)|^2}-2\pi\im Z\cdot\im\tau^{-1}\cdot\im Z\right]\cdot\frac{\nu^2}{2}\right]}
\ee
from which we see, given that $H(X_2)$ is modular invariant, that $\Phi(\tau)$ must carry modular weight $(1/2,1/2)$.

Finally, again following \cite{DHokerGreen}, the remaining factor of $\exp{\left[-\pi\text{Im}z\cdot\tau^{-1}\cdot\text{Im}z\right]}$ can be recast as a real-valued characteristic with (\ref{charrelation}).  This leads to an alternate expression:
\be
\label{Phi}
\Phi(\tau)=\exp{\left[\int_{T^4}d^4x\log{\left|\theta\genfrac[]{0pt}{0}{x'_1\text{ }x'_2}{x''_1\text{ }x''_2}(0|\tau)\right|^2}\right]}
\ee
where the integration is taken over a square, unit 4-torus.

\end{itemize}

\section{Degeneration}
\label{Degen}

We move to examining the behavior of the fundamental quantities associated with Riemann surfaces under degeneration.  From now on we take the original surface $X$ to have genus two.  Such a surface can degenerate to two tori with one node each connected by a long, thin tube, or to a single torus with two nodes and an attached handle.  The expressions derived in this section will be crucial for checking our later results against predictions from CFT.

\subsection{Basics of the Separating Degeneration}

In the case of the separating degeneration, we are imagining pinching a cycle of the Riemann surface which is trivial in homology, resulting in two connected surfaces with $g_1+g_2=g$.  The standard reference for a detailed explanation of the setup is \cite{Fay}, section III.  We include a more abridged description for our particular case of interest.

We take two tori, labeled $T_1^{(1)}$ and $T_1^{(2)}$, and remove a point from each, respectively $p_1$ and $p_2$.  In a neighborhood around each point we designate an annulus, and we will `glue together' the two annuli by identification.  We have a complex parameter $t$, for which $|t|$ will control the size of the annuli.  This construction forms a surface of genus two (more accurately a family of surfaces) which, as $|t|\to 0$, can be regarded as the two tori attached by a long, thin tube.

$t$ then appears in the quantities associated with the degenerating surface.  In the leading order limit, the diagonal entries of the period matrix simply become the structure constants of the two tori.  In general, the diagonal elements of $\tau$ have an even expansion in $t$, while the off-diagonals are odd.  In order to expand all of our desired quantities beyond leading order, we require the expansion of the period matrix up to order $t^3$.  This is given by \cite{MT}
\begin{multline}
\label{pmsepdegen}
\tau\to\left(\begin{matrix}\tau_1 & 0 \\ 0 & \tau_2\end{matrix}\right)+2\pi it\left(\begin{matrix}0 & 1 \\ 1 & 0\end{matrix}\right)+(2\pi it)^2\left(\begin{matrix}-2\partial\log{\eta(\tau_2)} & 0 \\ 0 & -2\partial\log{\eta(\tau_1)}\end{matrix}\right) \\
+(2\pi it)^3\left(\begin{matrix}0 & 4\partial\log{\eta(\tau_1)}\partial\log{\eta(\tau_2)} \\ 4\partial\log{\eta(\tau_1)}\partial\log{\eta(\tau_2)} & 0\end{matrix}\right).
\end{multline}

The one-forms associated to the surface likewise degenerate, in a manner compatible with (\ref{omegaperiods}) and (\ref{pmsepdegen}).  Since the calculations we undertake will depend only on the period matrix, we will not present the one-form relations.  Additionally, there is a second way in which our surface could degenerate -- we could allow a homologically nontrivial cycle to degenerate, which would leave us with a torus with an attached handle (called the non-separating degeneration).  However, since the separating degeneration will be sufficient to check our results, we will not go into particulars about the non-separating case.

\subsection{Theta Constants}

An easy first consequence of (\ref{pmsepdegen}) is the degeneration of the theta constants, which are determined entirely by the behavior of the period matrix.  First, we establish notation.  It will be useful to regard the sixteen genus two characteristics as coming from products of genus one characteristics.  This means that we can uniquely label a genus two theta constant by two genus one characteristics, e.g. we would write
\be
\theta_{14}(0|\tau)\equiv\theta\left[\begin{matrix}\frac{1}{2} & 0 \\ \frac{1}{2} &\frac{1}{2}\end{matrix}\right](0|\tau).
\ee
Using (\ref{pmsepdegen}) and (\ref{thetacharac}), we find that genus two theta constants degenerate to order $t^3$ as
\begin{multline}
\label{thetaconstantsepdegen}
\theta_{ij}(0|\tau)\to\theta_i\theta_j+t\theta'_i\theta'_j+\frac{(4\pi it)^2}{2}[\dot{\theta}_i\dot{\theta}_j-\partial\log{\eta(\tau_1)}\theta_i\dot{\theta}_j-\partial\log{\eta(\tau_2)}\dot{\theta}_i\theta_j] \\
+\frac{(4\pi i)^2t^3}{6}[\dot{\theta'}_i\dot{\theta'}_j-3\partial\log{\eta(\tau_1)}\theta'_i\dot{\theta'}_j-3\partial\log{\eta(\tau_2)}\dot{\theta'}_i\theta'_j+6\partial\log{\eta(\tau_1)}\partial\log{\eta(\tau_2)}\theta'_i\theta'_j].
\end{multline}
Here we have used a dot to indicate a $\tau$ derivative and a prime to indicate a $z$ derivative.  The arguments of the genus one theta constants are omitted to save space.

\subsection{$\chi_{10}$}

One modular form that appears frequently at genus two is Igusa's cusp form \cite{Igusa}, given by the product of the squares of the ten even theta constants:
\be
\label{chi10}
\chi_{10}(\tau)=\prod_{(\delta,\epsilon)\text{ even}}^{10}\theta^2\genfrac[]{0pt}{0}{\delta}{\epsilon}(0|\tau).
\ee
This is a modular form of weight (10,0).  Its modular invariant norm is given by
\be
\label{modchi10}
||\chi_{10}(\tau)||=2^{-12}(\det{\im\tau})^5|\chi_{10}(\tau)|.
\ee
From (\ref{thetaetarelation}) we see that there is a sense in which $\chi_{10}$ generalizes the eta function to genus two.  We are interested in knowing its behavior under degeneration.

First, it will be important to examine the ten even genus two characteristics.  We note that nine of them are of the form $\theta_{ee'}$ where $e$ and $e'$ are even genus one characteristics; we call these even-even constants.  The remaining one is $\theta_{11}$, which we call odd-odd.  From (\ref{thetaconstantsepdegen}), keeping in mind that a derivative in $z$ flips parity, we see that the even-even theta constants separate as
\be
\label{evenevensep}
\theta_{ee'}(0|\tau)\to \theta_{e}\theta_{e'}+\frac{(4\pi it)^2}{2}[\dot{\theta}_e\dot{\theta}_{e'}-\partial\log{\eta(\tau_1)}\theta_e\dot{\theta}_{e'}-\partial\log{\eta(\tau_2)}\dot{\theta}_e\theta_{e'}].
\ee
The odd-odd constant, however, will behave as
\begin{multline}
\theta_{11}(0|\tau)\to t\theta'_1\theta'_1+\frac{(4\pi i)^2t^3}{6}[\dot{\theta'}_1\dot{\theta'}_1-3\partial\log{\eta(\tau_1)}\theta'_1\dot{\theta'}_1\\
-3\partial\log{\eta(\tau_2)}\dot{\theta'}_1\theta'_1+6\partial\log{\eta(\tau_1)}\partial\log{\eta(\tau_2)}\theta'_1\theta'_1]
\end{multline}
Using the identity (\ref{thetaetarelation}), we can immediately simplify this to
\be
\theta_{11}(0|\tau)\to 4\pi^2t\eta^3(\tau_1)\eta^3(\tau_2)\left[1-\frac{1}{2}(4\pi it)^2\partial\log{\eta(\tau_1)}\partial\log{\eta(\tau_2)}\right].
\ee
Returning to the even-even piece, the leading term is simply products of even theta constants, so by (\ref{thetaconstantsepdegen}) it will be $2^6\eta^9(\tau_1)\eta^9(\tau_2)$.  At next order, we have nine terms, and in each we replace one of the $\theta_e\theta_{e'}$ with the second order term from (\ref{evenevensep}).  The result can be organized into terms that match the $\tau$ derivative of (\ref{thetaetarelation}), so that in the end everything is expressed in terms of eta and derivatives thereof.  In total we find
\be
\prod_{i,j=2}^4\theta_{ij}(0|\tau)\to 2^6\eta^9(\tau_1)\eta^9(\tau_2)\left[1-\frac{9}{2}(4\pi it)^2\partial\log{\eta(\tau_1)}\partial\log{\eta(\tau_2)}\right].
\ee
In total, then, we have the product of the even theta constants going to
\be
\prod_{(\delta,\epsilon)\text{ even}}\theta\genfrac[]{0pt}{0}{\delta}{\epsilon}(0|\tau)\to 2^8\pi^2t\eta^{12}(\tau_1)\eta^{12}(\tau_2)[1-5(4\pi it)^2\partial\log{\eta(\tau_1)}\partial\log{\eta(\tau_2)}],
\ee
and from this we have $\chi_{10}$ degenerating as
\be
\label{chi10degen}
\chi_{10}(\tau)\to 2^{16}\pi^4t^2\eta^{24}(\tau_1)\eta^{24}(\tau_2)[1-10(4\pi it)^2\partial\log{\eta(\tau_1)}\partial\log{\eta(\tau_2)}],
\ee
agreeing with \cite{Tuite}.

\subsection{$H(X_2)$ and $\Phi(\tau)$}

Next we examine the degenerating behavior of $H(X_2)$, defined in (\ref{H}).  We restate it here, specialized to genus two:
\be
\log{H(X_2)}=\int_{J(X_2)}\log{||\theta(Z|\tau)||}\cdot\frac{\nu^2}{2}.
\ee

This adds slightly more complication than $\chi_{10}$ due to the integration.  However, the fact that $H$ is defined over the Jacobian, which depends only on $\tau$, will keep things manageable.

To begin we examine the measure.  From the definition (\ref{nu}) of $\nu$ we have that
\be
\frac{\nu^2}{2}=\left(\frac{i}{2}\right)^2\frac{dZ_1\wedge d\bar{Z}_1\wedge dZ_2\wedge d\bar{Z}_2}{\det{\im\tau}}.
\ee
Using the parameterization (\ref{torusmu}) for each pair of coordinates, this simplifies to
\be
\frac{\nu^2}{2}=d\alpha_1d\beta_1d\alpha_2d\beta_2.
\ee
Expressed in these coordinates, the measure has the pleasant property of being $\tau$-independent, and thereby independent of the degeneration parameter $t$.  It remains to evaluate the $t$ series of the integrand at each order.

\subsubsection{Leading Order}

The theta function in the integrand behaves simply as $||\theta(Z|\tau)||\to||\theta(Z_1|\tau_1)||\text{ }||\theta(Z_2|\tau_2)||$.  Thus, using our result (\ref{intthetatorus}) for the integration of $||\theta||$ over a torus, we find the leading order term in the degeneration series to be
\begin{multline}
\log{H(X_2)}\to \int_{T_1^{(1)}\times T_1^{(2)}}\log{\left[||\theta(Z_1|\tau_1)||\text{ }||\theta(Z_2|\tau_2)||\right]} d\alpha_1d\beta_1d\alpha_2d\beta_2 \\ =\int_{T_1^{(1)}}\log{||\theta(Z_1|\tau_1)||}d\alpha_1d\beta_1+\int_{T_1^{(2)}}\log{||\theta(Z_2|\tau_2)||}d\alpha_2d\beta_2=\log{\left[||\eta(\tau_1)||\text{ }||\eta(\tau_2)||\right]}.
\end{multline}
From (\ref{PhiHrel}) we can quickly translate this into a result for $\Phi(\tau)$ by writing
\be
\label{PhiDegen}
\Phi(\tau)=\frac{H^2(X_2)}{\sqrt{\det{\im\tau}}}\to \frac{||\eta(\tau_1)||^2}{\sqrt{\im\tau_1}}\frac{||\eta(\tau_2)||^2}{\sqrt{\im\tau_2}}=|\eta(\tau_1)|^2|\eta(\tau_2)|^2,
\ee
which agrees with the calculations of \cite{DHokerGreen}.

\subsubsection{Subleading Order}

Moving beyond the leading order in $t$, we shift to looking at $\Phi(\tau)$.  From (\ref{PhiHrel}), we can write
\be
\Phi^{1/2}(\tau)=\exp{\left[\int\left[-\pi\left(\im Z\cdot\im\tau^{-1}\cdot\im Z\right)+\log{|\theta(Z|\tau)|}\right]d\alpha_1d\beta_1d\alpha_2d\beta_2\right]}.
\ee
Writing $Z=\alpha\tau+\beta$ and using the expansion (\ref{pmsepdegen}) of the period matrix, the term from the norm becomes
\begin{multline}
\label{normdegen}
\im Z\cdot\im\tau^{-1}\cdot\im Z\to \alpha_1^2\im\tau_1+\alpha_2^2\im\tau_2\\
-(4\pi it)^2\frac{\pi i}{4}[\alpha_1^2\partial\log{\eta(\tau_2)}+\alpha_2^2\partial\log{\eta(\tau_1)}]+(4\pi i\bar{t})^2\frac{\pi i}{4}[\alpha_1^2\partial\log{\eta(\bar{\tau}_2)}+\alpha_2^2\partial\log{\eta(\bar{\tau}_1)}].
\end{multline}
Expanding the theta function will yield terms of orders $t^0, t^1$ and $t^2$.  The terms of order $t$ all involve theta functions acted on by single $z$ derivatives.  Referring back to the calculations of section \ref{integralcalc}, recall that when the (log of) the theta function was cast in its product form (\ref{thetaprod}) and then integrated, only the first of three terms survived due to periodicity of the second and third terms in $\beta$.  This lone surviving term is killed by a $z$ derivative (and the periodicity of the other terms is untouched), so the order $t$ terms that arise from expanding the theta function become trivial upon integration.  With these terms removed, the holomorphic part of the log of the theta function goes to
\begin{multline}
\label{logthetadegen}
\frac{1}{2}\log{\theta(Z|\tau)}\to \frac{1}{2}\log{[\theta(Z_1|\tau_1)\theta(Z_2|\tau_2)]}\\
-\frac{1}{4}(4\pi it)^2\left[\partial\log{\eta(\tau_2)}\partial_\tau\log{\theta(Z_1|\tau_1)}+\partial\log{\eta(\tau_1)}\partial_\tau\log{\theta(Z_2|\tau_2)}\right],
\end{multline}
with an analogous antiholomorphic result.  We see that the terms from (\ref{normdegen}) and (\ref{logthetadegen}) are exactly of the form (\ref{dintthetatorus}) such that, upon integration and exponentiation, we have
\be
\label{PhiDegenSubleading}
\Phi^{1/2}(\tau)\to |\eta(\tau_1)||\eta(\tau_2)|-\frac{1}{4}(4\pi it)^2\partial\log{\eta(\tau_1)}\partial\log{\eta(\tau_2)}-\frac{1}{4}(4\pi i\bar{t})^2\partial\log{\eta(\bar{\tau}_1)}\partial\log{\eta(\bar{\tau}_2)}.
\ee

\section{Relation to the Free Boson CFT}
\label{CFT}

Having laid the groundwork, we are now ready to describe how the quantities of section \ref{AlgGeom} appear in CFT.  We saw in (\ref{zgdet}) that the free boson partition function is directly linked to the determinant of the scalar laplacian on the worldsheet Riemann surface $X_g$.  Faltings' invariant $\delta(X_g)$ can, as well, be written in terms of $\det{\Delta_0}$, linking it with the partition function -- the precise relation is \cite{DeJong}
\be
\label{zdrelation}
Z_g=\left[\frac{(\det{\Delta_0})_{X_g}}{V_{X_g}}\right]^{-\frac{1}{2}}=c(g)e^{\frac{\delta(X_g)}{12}}.
\ee
Here $c(g)$ is a genus-dependent normalization constant (independent of $\tau$), and we are implicitly choosing the Arakelov metric for our surface.  We postpone discussion of normalization until section \ref{zdegen}, where the usual CFT normalization at genus one will fix this constant.  By relating $\delta(X_g)$ to simpler invariants, we can begin to derive expressions for the partition function.

Let us quickly verify that this method produces the expected results on the torus.  There we see, from (\ref{dstrelation}) and (\ref{zdrelation}), that our partition function is simply proportional to a power of $T$:
\be
Z_1\sim T^{1/3}(X_1).
\ee
$T(X_1)$ is given, up to constants, by $(\text{Im}\tau)^{-3/2}|\eta(\tau)|^{-6}$ \cite{DeJong}, so we find the expected result
\be
\label{z1delta}
Z_1(\tau)= \frac{|\eta(\tau)|^{-2}}{\sqrt{\text{Im}\tau}}.
\ee

\subsection{$Z_2$}

At genus two we can again start from (\ref{dstrelation}), but now we have a nontrivial factor of $S$ appearing.  It will be more convenient for us to work with $H$ than $S$ for the non-factorizing terms.  Luckily, it turns out that $S(X_2)$ and $H(X_2)$ satisfy a relation of the form
\be
S(X_2)=||\chi_{10}(\tau)||^{-1/4}H(X_2)^4,
\ee
with $||\chi_{10}(\tau)||$ as given in (\ref{modchi10}).  $T(X_2)$ is proportional to $||\chi_{10}||^{-5/16}$ \cite{DeJong}, so after combining everything the exact relation is
\be
\delta(X_2)=-16\log{2\pi}+12\log{2}-\log{||\chi_{10}||}-4\log{H}(X_2)
\ee
which gives, up to normalization,
\be
\label{z2h}
Z_2=||\chi_{10}||^{-1/12}H(X_2)^{-1/3}.
\ee
Now, using (\ref{PhiHrel}) to rewrite $H(X_2)$ in terms of $\Phi(\tau)$, we find an expression for $Z_2$ as
\be
\label{z2}
Z_2(\tau)=\frac{2|\chi_{10}(\tau)|^{-1/12}\Phi(\tau)^{-1/6}}{\sqrt{\det{\text{Im}\tau}}}.
\ee

This expression is pleasing in that it mimics the genus one result (\ref{z1delta}) above.  However, there is subtlety hidden in the definition (\ref{Phi}) of $\Phi(\tau)$.  Though it may appear otherwise, $\Phi$ cannot be written as the holomorphic square of a function on moduli space \cite{DHokerGreen}.  This loss of holomorphic factorization is related to the conformal anomaly, parameterized by the theory's central charge $c$ \cite{BelavinKnizhnik}.  In fact, we have a specific expectation for the form of this obstruction -- one expects that the obstruction shows up at the level of $\det{\Delta_0}$, which factorizes as \cite{evthesis}
\be
\det{\Delta_0}\simeq e^{c S_L}|\det{\partial_0}|^2.
\ee
Here the nonfactorizing part is the exponential of a quantity known as the Liouville action $S_L$, which is given up to normalization by
\be
\label{sl}
S_L=\int_{X_g} d^2z\text{ }\log{g_{z\bar{z}}}\text{ }\partial_z\partial_{\bar{z}}\log{g_{z\bar{z}}}.
\ee
Our next task will be to show how such expressions show up in $Z_2$.

\subsection{Holomorphic Obstruction}
\label{holoobst}

For $g=2$ the expression (\ref{sl}), by the definition (\ref{arakelovmetric}) of the Arakelov metric, becomes (we have passed to integration over the (1,1)-form $\mu$ by replacing $dz^2$ with $idz\wedge d\bar{z}$)
\be
\label{thing1}
S_L=-4\pi\int_{X_2}\log{g_{z\bar{z}}}\cdot\mu(z).
\ee
Now consider (\ref{GEgRelation}).  If we apply $\int_{X_2^2}\mu(z)\mu(w)$ to each side, the left-hand side vanishes by the property (\ref{akgreenrel}) of $G(z,w)$.  We are left with
\be
\int_{X_2}\log{g_{z\bar{z}}}\cdot\mu(z)=-\log{\det{\im\tau}}-\int_{X_2^2}\log{||E(z,w)||^2}\cdot\mu(z)\mu(w),
\ee
and combining the previous two equations we find a fairly simple expression for the genus two Liouville action (in the Arakelov metric) as
\be
\label{sl3}
S_L=4\pi\log{\det{\im\tau}}+4\pi \int_{X_2^2}\log{||E(z,w)||^2}\cdot\mu(z)\mu(w).
\ee
This form will nicely facilitate us re-expressing $S_L$ in terms of the invariants of section \ref{Arakelov}.  We use the explicit expression (\ref{primeform}) for the prime form to write
\be
\log{||E(z,w)||}=\log{||\theta_\Delta(z-w)||}-\log{||h_\Delta(z)||}-\log{||h_\Delta(w)||}.
\ee
We recognize that (\ref{sl3}) can be written in terms of invariants as
\be
S_L=4\pi\log{\det{\im\tau}}-8\pi A(X_2)-16\pi\log{C(X_2)}.
\ee
$A(X_2)$ is related to $H(X_2)$ by \cite{Bost}
\be
\label{AHrelation}
A(X_2)=-\frac{1}{8}\log{||\chi_{10}||}+\frac{3}{2}\log{H(X_2)},
\ee
which lets us combine the equations above to express (\ref{z2h}) in terms of the Liouville action as
\be
\label{z2sl}
Z_2=\exp{\left[\frac{1}{36\pi}S_L\right]}(\det{\im\tau})^{1/9}||\chi_{10}||^{-1/9}C(X_2)^{4/9}.
\ee

\subsection{Degeneration of Partition Functions}
\label{zdegen}

We are now in a position to apply the results of section \ref{Degen} to the expressions we've derived.  Before doing so we will review the expected form of the results.  From general CFT arguments \cite{friedanshenker,evthesis}, we would expect the genus two partition function (vacuum correlation function) to behave under a separating degeneration as
\be
\label{separatingschematic}
Z\to\sum_{h_i,\bar{h}_j}t^{h_i}\bar{t}^{\bar{h_j}}\braket{\Or_i(0)}_{\tau_1}\braket{\Or_j(0)}_{\tau_2},
\ee
where the coefficients are one-point functions of operators $\Or_i$ with weights $h_i,\bar{h}_i$.

\subsubsection{Noncompact Free Boson -- Leading Order}

At the bottom of the spectrum there is a unique field with $h=\bar{h}=0$ called the vacuum.  Its multipoint functions with itself all give the partition function, so to leading order at genus two the separating degeneration should have the form
\be
Z_2(\tau)\to Z_1(\tau_1)Z_1(\tau_2).
\ee
However in practice there is more to this story, because we expect leading-order divergences in the degeneration parameter, owing to the conformal anomaly \cite{friedanshenker}\cite{NelsonLectures}.  At this point we note that the degeneration parameter $t$, as we've defined it, must be of non-zero modular weight.  This can be seen, for instance, in the degeneration series (\ref{pmsepdegen}) of the period matrix.  In order for the modular weights of the various terms to match, $t$ must be of weight (-1,0).  In order to maintain modular invariance in the leading term, then, we should expect that the anomalous factors of $t$ appear with appropriate coefficients.  According to Wentworth \cite{Wentworth} the modular invariant parameter we should consider is given, for the separating degeneration, by
\be
||t||=t\sqrt{g_{z_1\bar{z}_1}(0)}\sqrt{g_{z_2\bar{z}_2}(0)}.
\ee
For a genus one surface with the Arakelov metric we have $g_{z\bar{z}}(0)=4\pi^2\eta^4(\tau)$, so for our setup of a degenerating genus two surface we should take $||t||=4\pi^2 t\eta^2(\tau_1)\eta^2(\tau_2)$.

With this in mind we examine the leading-order degeneration of $Z_2(\tau)$ as given in (\ref{z2}).  The degenerations of $\chi_{10}(\tau)$ and $\Phi(\tau)$ were given in (\ref{chi10degen}) and (\ref{PhiDegen}), respectively.  $\det{\im\tau}$ simply behaves as $\det{\im\tau}\to\im\tau_1\im\tau_2$, and so we have
\be
Z_2(\tau)\to|4\pi^2t\eta^2(\tau_1)\eta^2(\tau_2)|^{-1/6}\frac{|\eta(\tau_1)|^{-2}}{\sqrt{\im\tau_1}}\frac{|\eta(\tau_2)|^{-2}}{\sqrt{\im\tau_2}}=||t||^{-1/6}Z_1(\tau_1)Z_1(\tau_2).
\ee
We have obtained the normalized genus one partition functions with no extraneous multiplicative constants -- this will serve as our normalization criterion for the genus two partition function, fixing the constant in (\ref{zdrelation}).

\subsubsection{The Compact Free Boson -- Subleading Order}

In order to derive a simple expected form for the subleading terms, we move to examining the compact free boson.  This is done by imposing the identification $\varphi\sim\varphi+2\pi R$, causing the partition function to be multiplied by an additional $R$-dependent term.  The necessity of its appearance can be seen by noting that the path integral (\ref{zgpathint}) now breaks into topological sectors.  Physically, the boson is acquiring momentum and winding modes along the compact direction.  The form of this new term will be
\be
\label{freebosonmomentumlattice}
\sqrt{\det{\im\tau}}\text{ }Z_g^{\text{M.L.}}=\sqrt{\det{\im\tau}}\sum_{p,\bar{p}\in\Gamma_g}\exp{\left[\frac{2\pi i}{4}(p\cdot\tau\cdot p-\bar{p}\cdot\bar{\tau}\cdot\bar{p})\right]}
\ee
where $p$ and $\bar{p}$ are $g$-component objects which take values in the momentum lattice
\be
\Gamma_g=\left(\frac{x_i}{R}+y_iR,\frac{x_i}{R}-y_iR\bigg|(x_i,y_i)\in\Z^2\right).
\ee
The expression (\ref{freebosonmomentumlattice}) can be thought of as a generalized theta function defined on $\Gamma_g$.  The total partition function for the compact free boson is then
\be
\mathcal{Z}_g=Z_g^{\text{Pre}}Z^{\text{M.L.}}_g
\ee
where the prefactor $Z_g^{\text{Pre}}$ is the noncompact free boson partition function which we've been examining up until now, multiplied by $\sqrt{\det{\im\tau}}$.  For example, on the torus we have the well-known result
\be
\mathcal{Z}_1(\tau)=|\eta(\tau)|^{-2}\sum_{x,y\in\Z^2}\exp{\left[\frac{2\pi i}{4}\left[\tau\left(\frac{x}{R}+yR\right)^2-\bar{\tau}\left(\frac{x}{R}-yR\right)^2\right]\right]}.
\ee
Returning to genus two, our result (\ref{z2}) tells us that $Z_2^{\text{Pre}}=2|\chi_{10}|^{-1/12}\Phi^{-1/6}$, so the genus two compact free boson partition function is
\be
\label{z2compact}
\mathcal{Z}_2(\tau)=2|\chi_{10}(\tau)|^{-1/12}\Phi^{-1/6}(\tau)\sum_{p,\bar{p}\in\Gamma_2}\exp{\left[\frac{2\pi i}{4}(p\cdot\tau\cdot p-\bar{p}\cdot\bar{\tau}\cdot\bar{p})\right]}.
\ee

Using the period matrix's degeneration  (\ref{pmsepdegen}) it is straightforward to expand $Z_2^{\text{M.L.}}$ to second order in $t$.  The leading term simply becomes the product of the two genus one lattice sums -- to second order, then, we find the result
\begin{multline}
Z_2^{\text{M.L.}}(\tau)\to\bigg[1+\frac{1}{2}(4\pi it)^2(\partial_{\tau_1}\partial_{\tau_2}-\partial\log{\eta(\tau_2)}\partial_{\tau_1}-\partial\log{\eta(\tau_1)}\partial_{\tau_2})\\
+\frac{1}{2}(4\pi i\bar{t})^2(\partial_{\bar{\tau}_1}\partial_{\bar{\tau}_2}-\partial\log{\eta(\bar{\tau}_2)}\partial_{\bar{\tau}_1}-\partial\log{\eta(\bar{\tau}_1)}\partial_{\bar{\tau}_2})\bigg]Z_1^{\text{M.L.}}(\tau_1)Z_1^{\text{M.L.}}(\tau_2)
\end{multline}
acting on the product of genus one momentum lattices.  The remaining contribution comes from $Z_2^{\text{Pre}}$.  Using the results (\ref{chi10degen}) and (\ref{PhiDegenSubleading}) for the order $t^2$ degenerations of $\chi_{10}$ and $\Phi$, we find that
\begin{multline}
Z_2^{\text{Pre}}\to ||t||^{-1/6}|\eta(\tau_1)|^{-2}|\eta(\tau_2)|^{-2}\\
[1+\frac{1}{2}(4\pi it)^2\partial\log{\eta(\tau_1)}\partial\log{\eta(\tau_2)}+\frac{1}{2}(4\pi i\bar{t})^2\partial\log{\eta(\bar{\tau}_1)}\partial\log{\eta(\bar{\tau}_2)}].
\end{multline}
Combining both the prefactor and momentum lattice terms, we find that our expression (\ref{z2compact}) for the genus two compact free boson degenerates up to order $t^2$ as
\begin{multline}
\label{z2compactorder2degen}
\mathcal{Z}_2(\tau)\to ||t||^{-1/6}Z_1^{\text{Pre}}(\tau_1)Z_1^{\text{Pre}}(\tau_2)[1+\frac{1}{2}(4\pi it)^2(\partial_{\tau_1}-\partial\log{\eta(\tau_1)})(\partial_{\tau_2}-\partial\log{\eta(\tau_2)})\\
+\frac{1}{2}(4\pi i\bar{t})^2(\partial_{\bar{\tau}_1}-\partial\log{\eta(\bar{\tau}_1)})(\partial_{\bar{\tau}_2}-\partial\log{\eta(\bar{\tau}_2)})]Z_1^{\text{M.L.}}(\tau_1)Z_1^{\text{M.L.}}(\tau_2).
\end{multline}

\subsubsection{Calculation from CFT}

Now we compute the expected form of the $\mathcal{Z}_2$ degeneration using only input from CFT.  At a generic radius, the first contribution to (\ref{separatingschematic}) we would expect comes from the stress tensor $T$, which has $h=2, \bar{h}=0$, and its antiholomorphic counterpart $\bar{T}$ which has $h=0, \bar{h}=2$.  This means we should see terms proportional to $t^2$ and $\bar{t}^2$, with nothing appearing at order $t$ and no mixed $t\bar{t}$ term.

We can check that this is giving us the correct correlation function.  On the torus, the one-point function of the (normalized) stress tensor has the form \cite{YellowBook}
\be
\braket{T}_\tau=2\sqrt{2}\pi i\partial_\tau \mathcal{Z}_{1}=2\sqrt{2}\pi i\partial_\tau[Z^{\text{Pre}}_1(\tau)Z^{\text{M.L.}}_1(\tau)].
\ee
Using $Z^{\text{Pre}}_1=|\eta(\tau)|^{-2}$, we have
\be
\label{stresstensoronepoint}
\braket{T}_\tau=2\sqrt{2}\pi iZ_1^{\text{Pre}}(\tau)[\partial_\tau-\partial\log{\eta(\tau)}]Z^{\text{M.L.}}_1(\tau).
\ee
Writing (\ref{separatingschematic}) out to second order, then, our expectation for the degeneration series of the full compact free boson partition function is
\be
\label{z2secondorder}
\mathcal{Z}_2\to \mathcal{Z}_1(\tau_1)\mathcal{Z}_1(\tau_2)+t^2\braket{T}_{\tau_1}\braket{T}_{\tau_2}+\bar{t}^2\braket{\bar{T}}_{\bar{\tau}_1}\braket{\bar{T}}_{\bar{\tau}_2}.
\ee
Plugging (\ref{stresstensoronepoint}) into (\ref{z2secondorder}), we obtain
\begin{multline}
\mathcal{Z}_2(\tau)\to Z_1^{\text{Pre}}(\tau_1)Z_1^{\text{Pre}}(\tau_2)[1+\frac{1}{2}(4\pi it)^2(\partial_{\tau_1}-\partial\log{\eta(\tau_1)})(\partial_{\tau_2}-\partial\log{\eta(\tau_2)})\\
+\frac{1}{2}(4\pi i\bar{t})^2(\partial_{\bar{\tau}_1}-\partial\log{\eta(\bar{\tau}_1)})(\partial_{\bar{\tau}_2}-\partial\log{\eta(\bar{\tau}_2)})]Z_1^{\text{M.L.}}(\tau_1)Z_1^{\text{M.L.}}(\tau_2)
\end{multline}
which exactly matches (\ref{z2compactorder2degen}) up to the divergent factor of $||t||$.

\section{Conclusion}
\label{conclusion}

Our endeavor to write $Z_2(\tau)$ in terms of modular forms resulted in the expression
\be
\label{z2final}
Z_2(\tau)=\frac{2|\chi_{10}(\tau)|^{-1/12}\Phi(\tau)^{-1/6}}{\sqrt{\det{\text{Im}\tau}}},
\ee
with $\chi_{10}$ defined in (\ref{chi10}), $\Phi$ in (\ref{PhiHrel}), and our genus two surface carrying the Arakelov metric.  We checked that this expression satisfies the expected properties of holomorphic obstruction and degeneration.  The nonfactorizing part was expected to come in the form of a Liouville action, and section \ref{holoobst} showed that (\ref{z2final}) can be written in such a form.  When the genus two worldsheet degenerates, the coefficients of the series expansion in the degeneration parameter are expected to give correlation functions in the corresponding conformal field theory.  The degeneration calculations of section \ref{Degen} -- which we were able to take beyond the standard leading order -- were shown in section \ref{zdegen} to match exactly the CFT predictions, up to the expected leading order divergence.

The presence of $\chi_{10}$ in (\ref{z2final}), as well as the power of -1/12, should come as no surprise; when the partition function for the bosonic string (which in light-cone gauge involves 24 copies of the free boson) is expressed in terms of modular forms, it is simply proportional to $|\chi_{10}|^{-2}$ \cite{TwoLoopIV}.  That result, however, includes ghost contributions which are needed to cancel the total conformal anomaly.  Our expression feels the anomaly's effects through $\Phi$ and its inability to be written as a holomorphic square \cite{DHokerGreen}.

As mentioned in the introduction, aside from filling a conceptual gap in the literature, these results should prove useful to bootstrap investigations at higher genus.  Given that genus two worked out, one might wonder whether the corresponding expressions for $g>2$ could be simplified in a similar manner.  One challenge is that many of the simplifying relations employed here are valid only for hyperelliptic Riemann surfaces -- this includes all surfaces of genus one and two \cite{dejongthesis}.  Above genus two, however, not all surfaces are of this form, and the increased variety leads to decreased control over general expressions.  The derivation of simple expressions along the lines of (\ref{z2final}) for $g>2$ will therefore remain open to investigation.

\section*{Acknowledgments}

The author is deeply thankful to D. Robbins and O. Lunin for support, suggestions and feedback on this project.

\bibliographystyle{ieeetr}
\bibliography{DegenerationPaper}
	
\end{document}